\documentclass[twoside]{article}
\usepackage{times}
\usepackage[latin1]{inputenc}
\usepackage{t1enc}
\usepackage{a4}
\usepackage{epsfig}
\usepackage{graphicx}

\def\title{\noindent \Large \bf }
\def\author(s){\large \sc
\vspace*{4mm}\hspace*{11.6mm}\parbox[t]{158mm}}
\def\institute(s){\normalsize \rm
\vspace*{2mm}\hspace*{11.6mm}\parbox[t]{158mm}}
\def\text{\normalsize \rm
\vspace*{5mm}\hspace*{11.6mm}\parbox[t]{158mm}}

\begin{document}




\title{Search for and investigation of new stellar clusters using the data from
huge stellar catalogues}

\author(s){
Sergey Koposov$^{1,2}$, Elena Glushkova$^1$ \& Ivan Zolotukhin$^1$}

\institute(s){
$^1$
Sternberg Astronomical Institute, Universitetskiy pr., 13, 119992 Moscow,
Russia\\
$^2$
Max Planck Institute for Astronomy, Koenigstuhl 17, D-69117 Heidelberg,
Germany\\
math@sai.msu.ru
}

\text{
In this paper we  present several new methods of search and investigation
of stellar clusters using the data available from new huge stellar
catalogues (like 2MASS, USNO etc.) 
\begin{enumerate}
\item{} We have developed a new method of effectively searching star
catalogues (like USNO, 2MASS etc.) for stellar clusters of different
radii. This task is rather complex, and we do not know any successful
attempts to automate such procedure applied for the search of open clusters. 
This method employs a modified technique of detection using the convolution
with density functions.

\item{} Also, we have developed a rather robust method, which can be used to
determine whether an observed density peak is just an occasional
overdensity of field stars, or this is a real group of
evolutionary-related (lying on one isochrone) stars. That method is
capable of simultaneous finding the position of the isochrone of the
cluster. These last steps are mainly based on the fact that in real
clusters, only stars lying on the isochrone show a density peak, whereas
the field stars should demonstrate flat distribution. This fact
allows us to find the position of the isochrone of the cluster even when
the CMD is "noised" by field stars. So briefly, we can automatically
find out prospective candidate for clusters, confirm them, and determine
main parameters (age, radius, distance and color excess) of these clusters.
\end{enumerate}

We formalized and programmed our algorithms, and almost completed their
automation. Algorithms were thoroughly tested. The tests were performed
for several fields from 2MASS, which is our primary target catalogue,
and their results are quite promising. From a very
preliminary analysis of the 150 sq. degrees in the Galaxy Anticenter
region, we have found more than 10 NEW clusters (see one example on Fig. 1)
and determined their
parameters. In that region, we were also able to determine the
parameters for several well-known clusters. A set of new clusters
was found when we just looked at the region of Perseus arm using our
algorithm. It should be also noticed that the clusters, which we have 
discovered, are not just infrared clusters; most of them are clearly 
visible in the optical wavelength range.

Conclusion: As a final result of our work, we plan to obtain a new
catalogue of confirmed stellar clusters, containing several percent of
new clusters with homogeneously measured parameters. We believe it would
be interesting to try to apply these methods to several other catalogues
(DENIS, SDSS etc.).
}
\begin{figure}[!h]
\begin{center}
\includegraphics[height=7cm]{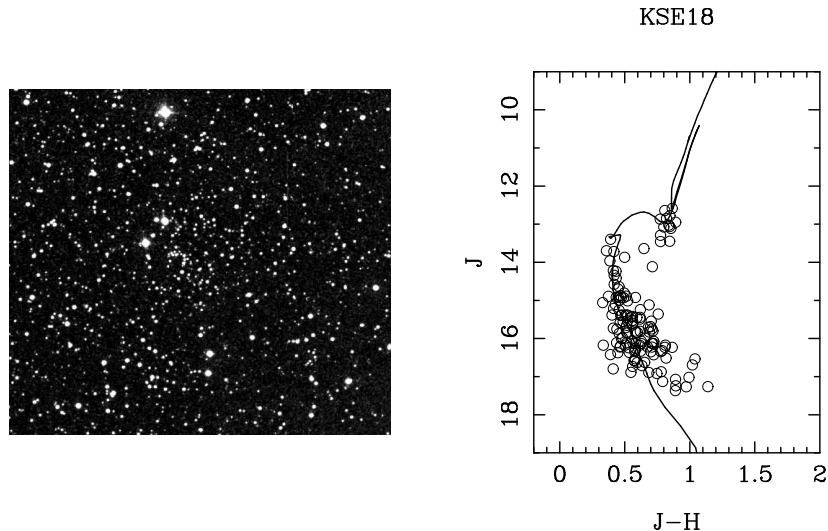}
\caption{The DSS image of one cluster discovered by us and the 2MASS color-magnitude
diagram for that cluster}
\end{center}
\end{figure}

\vfill
\end{document}